\documentclass[floatfix,pre,twocolumn,superscriptaddress]{revtex4}
\usepackage{tipa}
\pdfoutput=1
\usepackage{graphicx}
\usepackage[caption=false]{subfig}
\usepackage{alltt}
\usepackage{wrapfig}
\usepackage{verbatim}
\usepackage{alltt}
\usepackage{textcomp}
\usepackage{amsmath}
\usepackage{amsfonts}
\usepackage{bm}

\newcommand\ii{\mathrm{i}}

\newcommand\upd{\mathrm{d}}
\newcommand\vect{\boldsymbol}
\newcommand\tens{\boldsymbol}

\begin{document}

\title{The many-body reciprocal theorem and swimmer hydrodynamics}

\author{Dario Papavassiliou}
\email[]{d.papavassiliou@warwick.ac.uk}
\affiliation{Centre for Complexity Science and Department of Physics, University of Warwick, Coventry, CV7 4AL, UK}

\author{Gareth P. Alexander}
\affiliation{Centre for Complexity Science and Department of Physics, University of Warwick, Coventry, CV7 4AL, UK}

\begin{abstract}
We present a reinterpretation and extension of the reciprocal theorem for swimmers, extending its application from the motion of a single swimmer in an unbounded domain to the general setting, giving results for both swimmer interactions and general hydrodynamics. We illustrate the method for a squirmer near a planar surface, recovering standard literature results and extending them to a general squirming set, to motion in the presence of a ciliated surface, and expressions for the flow field throughout the domain. Finally, we present exact results for the hydrodynamics in two dimensions which shed light on the near-field behaviour.
\end{abstract}

\maketitle

\section{Introduction}
Microorganisms and self-propelled particles move and interact in a fluid environment dominated by the effects of viscosity~\cite{taylor1951,lighthill1976,lauga2009}. The long-range nature of Stokes flows suggest that hydrodynamics will have a significant effect on microorganism motion and behaviour. Examples of this are seen in the circular motion of {\it E. coli} near boundaries~\cite{diluzio2005,lauga2006}, attraction to surfaces~\cite{berke2008,kantsler2013}, synchronisation of cilia and flagella~\cite{goldstein2009,uchida2010}, interactions and scattering of pairs of swimmers~\cite{ishikawa2006,drescher2009}, and collective dynamics of suspensions~\cite{dombrowski2004,ishikawa2007,ramaswamy2010}.
The difficulty in determining swimmer hydrodynamics theoretically comes because the motion of the organism, which we wish to determine, constitutes a boundary condition. An elegant method for overcoming this was introduced by Stone \& Samuel~\cite{stone1996}, based on the Lorentz reciprocal theorem, a general relationship between two Stokes flows in the same domain but with different boundary conditions. This approach was originally developed as a way of determining the motion of a single swimmer, with a significant feature that it bypasses the need to calculate the full hydrodynamics, and in this capacity has become a standard tool in the active matter literature~\cite{squires2004,golestanian2007,crowdy2011,masoud2014,lauga2014}.

In fact, the reciprocal theorem may be used to determine essentially all aspects of the hydrodynamics of any swimmer problem, a point apparently not emphasised in the literature. Here we describe how this may be done and provide illustrations for a wide range of problems in two and three dimensions. We recover standard results in three dimensions for the interaction of a swimmer with a planar wall~\cite{spagnolie2012} and extend to the case of a swimmer interacting with an active carpet of cilia, showing that the surface activity may cause trapping or enhance deflection depending on the direction of the surface flow. 

The full utility of the reciprocal theorem is that problems in swimmer hydrodynamics may be solved easily if a conjugate Stokes drag solution is known. We illustrate this in two dimensions giving exact results for problems covering the interactions of two circular squirmers, or the motion of a single particle either inside or outside a circular domain, or swimming close to a planar boundary. 

\section{The reciprocal theorem}

The reciprocal theorem~\cite{happel1983} relates two solutions of the Stokes equations in the same domain, $D$, but with different boundary conditions. The solutions, $(\vect{u},\tens{\sigma})$ and $(\tilde{\vect{u}},\tilde{\tens{\sigma}})$, are related by the integral relation 
\begin{equation}
\int_{\partial D} \vect{u}\cdot \tilde{\tens{\sigma}} \cdot \hat{\vect{n}} = \int_{\partial D} \tilde{\vect{u}}\cdot \tens{\sigma} \cdot \hat{\vect{n}},
\label{eq:reciprocal_thm}
\end{equation} 
where $\hat{\vect{n}}$ is the unit normal pointing into the fluid, $\vect{u}$ is the velocity and $\tens{\sigma}$ is the stress. For a collection of swimmers moving with no force or torque at translational and rotational speeds $\vect{U}_i$ and $\vect{\Omega}_i$ due to surface slip velocities $\vect{u}_{s \, i}$, this splits as a sum over the boundaries, $S_i$, of each of the swimmers to give 
\begin{equation}
\sum_i \Bigl[ \vect{U}_i \cdot \tilde{\vect{\Phi}}_i + \vect{\Omega}_i \cdot \tilde{\vect{T}}_i \Bigr] = - \sum_i \int_{S_i} \vect{u}_{s \, i} \cdot \tilde{\tens{\sigma}} \cdot \hat{\vect{n}}_i\, ,
\label{eq:reciprocal} 
\end{equation}
where $\tilde{\vect{\Phi}}_i$ and $\tilde{\vect{T}}_i$ are the force and torque on the particle in a conjugate solution $(\tilde{\vect{u}},\tilde{\tens{\sigma}})$. By an appropriate choice of forces and torques in the conjugate problem each component of the swimmer motion can be obtained. The right-hand-side expresses this swimmer motion in terms of the boundary data, the slip velocities $\vect{u}_{s\,i}$, with the stress tensor of a conjugate Stokes drag problem playing the role of an integration kernel.

Although only a mild generalisation of the result given in~\cite{stone1996}, eq.~\eqref{eq:reciprocal} is tremendously general and allows more or less all aspects of swimmer hydrodynamics to be solved for. Loosely, this is as expected. The flow is given by the boundary data; the role of the reciprocal theorem is to identify the appropriate integration kernel.

\begin{figure}[t]
\includegraphics[width=0.48\textwidth]{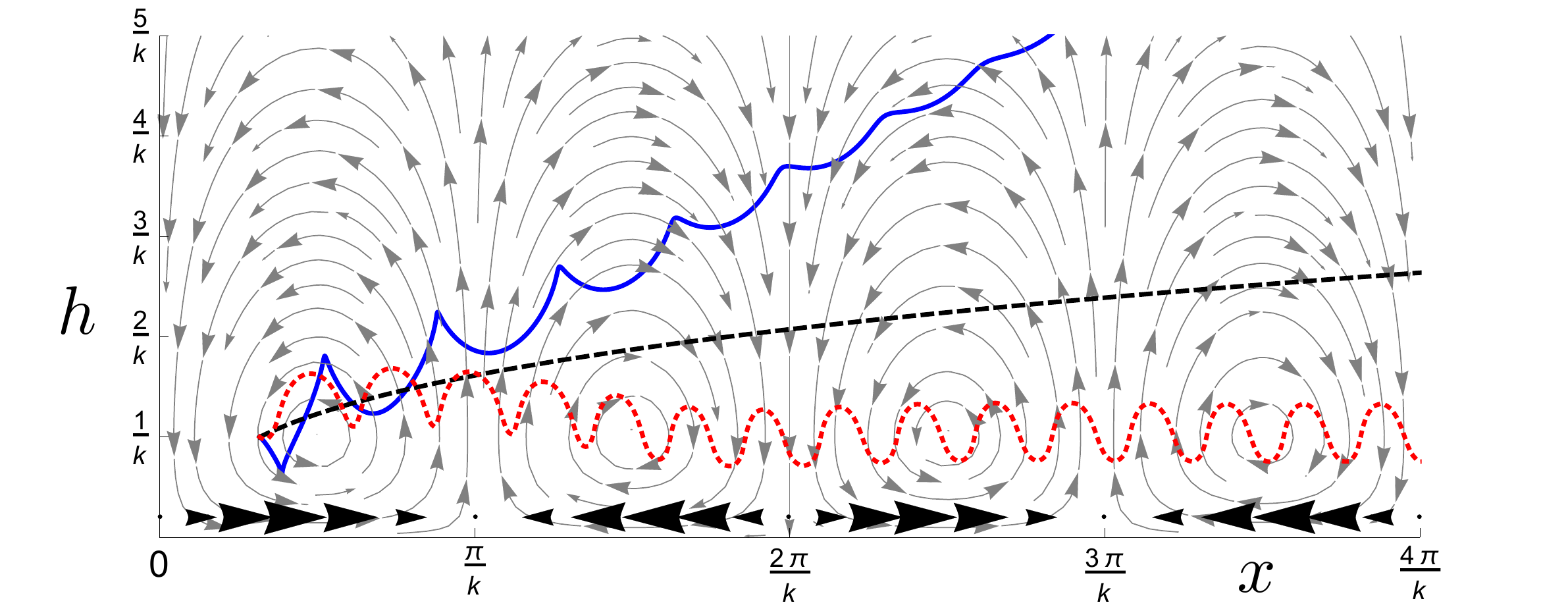}
\caption{The effect of a wall on swimmer trajectories. Close to a no-slip wall, a squirmer oriented parallel to the wall is repelled, following the black trajectory. This repulsion may be suppressed by a carpet of cilia on the wall generating a metachronal wave moving in the same direction, which can trap the swimmer in an advected orbit, shown by the red trajectory. If the metachronal wave opposes the swimming, the repulsion is enhanced, shown in blue. The black arrows represent the activity at $t=0$, resulting in the flow field, which decays exponentially with distance, shown in grey.
}
\label{fig0b}
\end{figure}

We illustrate this first for the case of a swimmer close to a planar boundary. The integration kernel is the stress tensor $\tilde{\tens{\sigma}}$ for the Stokes drag of a particle in the half-space, for which an approximate expression can be given in terms of fundamental flow singularities and the method of images. Retaining only the leading order contribution of a point force and torque in the half-space~\cite{blake1974}, the influence of the boundary on the motion of a swimmer at a distance $h \hat{\vect{z}}$ from the surface is
\begin{align}
\begin{gathered}
\vect{U}= \vect{U}^{\text{free}}-\frac{3}{8h^2}\frac{1}{4\pi} \int_{\textrm{swimmer}} \bigl[ u_{s\,z}\hat{\vect{n}}
+\vect{u}_{s}\hat{n}_{z}\\
+\hat{\vect{z}}(\vect{u}_{s} \cdot \hat{\vect{n}}+u_{s\,z}\hat{n}_{z}) \bigr]+\mathcal{O}(h^{-3}),
\end{gathered}
\label{approx_u}
\\
\begin{gathered}
\vect{\Omega} =\vect{\Omega}^{\text{free}} -\frac{3}{8 h^3}\frac{1}{8 \pi}\int_{\textrm{swimmer}}   \bigl[ (\hat{\vect{z}} \times \vect{u}_s)\hat{n}_z \\
+(\hat{\vect{z}} \times \hat{\vect{n}})u_{s\,z} \bigr]+\mathcal{O}(h^{-4}),
\end{gathered}
\label{approx_w}
\end{align}
where $\vect{U}^{\text{free}}$ and $\vect{\Omega}^{\text{free}}$ describe the swimmer's free motion in an unbounded domain. For an axisymmetric spherical squirmer of radius $a$ with a slip velocity $\vect{u}_s(\theta)=\sum_n A_n P_n (\theta) \hat{\vect{r}} +\sum_n B_n V_n(\theta)  \hat{\vect{\theta}} +\sum_n C_n V_n (\theta) \hat{\vect{\phi}}$, as defined in~\cite{lighthill1952}, this leading-order contribution to the motion due to this interaction is
\begin{equation}
\begin{gathered}
U_z-U_z^{\text{free}}=-\frac{3a^2}{16 h^2} \biggl[A_0 + (1+3 \cos 2\theta_0) \frac{(A_2-B_2)}{5} \biggr], \\
U_x-U_x^{\text{free}} =\frac{3a^2}{8 h^2} \sin 2\theta_0 \frac{(A_2-B_2)}{5}, \\
\Omega_y-\Omega_y^{\text{free}} =\frac{3a^2}{16 h^3} \sin 2\theta_0 \frac{(A_2-B_2)}{5},
\end{gathered}
\label{lauga_results}
\end{equation}
where $\theta_0$ is the angle the swimmer's head makes with the wall normal and $\hat{\vect{x}}$ is the direction of the head projected on the wall. All other components of the motion enter at higher order. Of these, particularly interesting is the rotation about the wall normal $\Omega_{z}$, which leads to circular trajectories of particles swimming parallel to the wall. When $\theta_0=\pi/2$ this rotation is given by
\begin{equation}
\Omega_{z}=\frac{a^3 C_2}{160h^4}+\mathcal{O}(h^{-5}),
\label{lauga_results2}
\end{equation}
where $C_2$ is the coefficient of the azimuthal slip velocity with angular dependence $\sin 2\theta$. Such a flow circulates in an easterly sense in the northern hemisphere but in a westerly sense in the southern hemisphere, so that a swimmer with this squirming mode qualitatively resembles bacteria like {\it E. coli} which have a counter-rotating head and tail. Indeed, such bacteria are known to swim in circles close to boundaries~\cite{diluzio2005}. The direction of the circling is set by the direction of the head-tail rotation and is right-handed if $C_2$ is positive. The radius of curvature is given by $160U^{\text{free}} h^4/a^3C_2$, indicating that this effect is strongly localised at boundaries, as reported experimentally~\cite{lauga2006}.

The interaction of swimmers with planar surfaces has been studied using a variety of techniques, such as boundary integrals~\cite{ishimoto2013} or singularity approximations of the swimming action~\cite{lauga2006,spagnolie2012}. The present approach using the reciprocal theorem reproduces all previous results, but it does so in a complementary fashion; the approximation is in the integration kernel, and the description of the swimmer motion remains exact. Eqs.~(\ref{approx_u})-(\ref{approx_w}) are the leading order interactions for an arbitrary slip velocity, with no restriction on axisymmetry or shape. Successive improvements can be given by using more accurate forms of the stress tensor in the conjugate problem. 

As a second illustration, using exactly the same integration kernel we can also find the motion of a passive particle propelled by the flow created by an active planar surface. For instance, this might correspond to transport by a carpet of cilia~\cite{blake1974b,satir2007}, for which a simple appropriate slip velocity is the metachronal wave $\vect{u}_s=U_0\sin(k x-\omega t)\hat{\vect{x}}$. Since the particle is no-slip, the reciprocal theorem reads $\vect{U}\cdot \tilde{\vect{\Phi}}+\vect{\Omega}\cdot \tilde{\vect{T}} = - \int_{\textrm{plane}} U_0\sin(k x-\omega t)\hat{\vect{x}} \cdot \tilde{\tens{\sigma}} \cdot \hat{\vect{z}}$,
so that its instantaneous velocity is given by
\begin{equation}
\begin{gathered}
U_x=U_0\mathrm{e}^{-k h} (1-k h)\sin(kx-\omega t), \\
U_z=-U_0\mathrm{e}^{-k h}  k h\cos(k x-\omega t),\\
\Omega_y=\tfrac{1}{2}U_0\mathrm{e}^{-k h}  k \sin(k x-\omega t).
\end{gathered}
\label{activewall}
\end{equation}
The effect of the surface is exponentially confined to a layer of a thickness comparable to the wavelength of the metachronal wave.The motion of a swimmer close to such an active wall is given by the linear superposition of eqs.~(\ref{approx_u}), (\ref{approx_w}) and (\ref{activewall}), the boundary components of eq.~(\ref{eq:reciprocal}). Depending on the wave's direction, the deflection of a squirmer parallel to a wall~\cite{spagnolie2012} may be enhanced or suppressed, as illustrated in fig.~(\ref{fig0b}).
 
The velocity (\ref{activewall}) is precisely the flow field produced by a waving sheet, as found by G.\ I.\ Taylor in his seminal analysis~\cite{taylor1951}. This makes sense because a small, spherical particle acts as a tracer which measures the local flow. Indeed, this simple example shows that the reciprocal theorem should be viewed as a solution for the entire swimmer hydrodynamics, and not just the motion. The fluid velocity at a point $\vect{x}$ is given by
\begin{equation}
\vect{u}(\vect{x}) \cdot \tilde{\vect{\Phi}} = - \sum_i \int_{S_i} \vect{u}_{s \, i} \cdot \tilde{\tens{\sigma}} \cdot \hat{\vect{n}} ,
\label{recipthm_flow}
\end{equation}
where $\tilde{\tens{\sigma}}$ is the stress tensor for the conjugate problem where the particles $S_i$ are all force and torque free and a point force $\tilde{\vect{\Phi}}$ is applied at $\vect{x}$. This greatly extends the scope of the reciprocal theorem from what was originally envisaged; not only does it give the swimmer motion, but a parallel analysis yields the full hydrodynamics.

This use of the reciprocal theorem allows any swimmer problem to be solved, provided the appropriate conjugate stress tensor is known. Fortunately many exact and approximate solutions of Stokes flow problems for a variety of geometries exist in the literature, so that the relevant stress tensor is available in many cases. As we have seen, the general three-dimensional case is easily solved using an approximate integration kernel, giving good asymptotic results. However, this approximation is not appropriate in the near-field, where the separation becomes comparable to the swimmer size. We will now use a classic two-dimensional exact result for the stress tensor to shed light on these cases.

\section{Integration kernel in two dimensions}

\begin{figure}[t]
\includegraphics[width=0.48\textwidth]{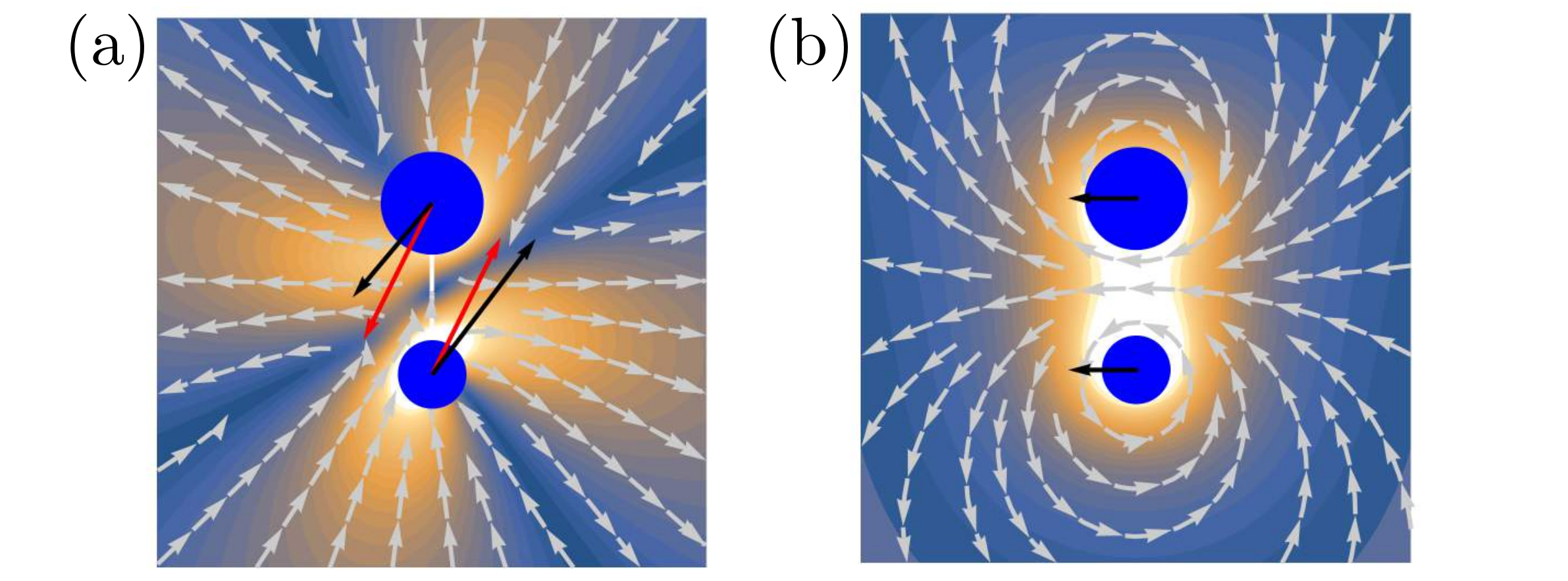}
        \caption{Flow field of the Stokes drag problem for two discs of unequal size. (a) Equal and opposite forces in the direction of the red arrows, (b) equal and opposite torques, clockwise on the upper disc. The black arrows show direction and relative magnitude of translational velocities and colour denotes flow speed in dimensionless units.
        }
        \label{fig3}
\end{figure}

In two dimensions an exact solution for the Stokes drag on two discs is available, in the case where there are equal and opposite forces $\tilde{\Phi}$ and equal and opposite torques $\tilde{T}$ acting on the discs~\cite{jeffery1921,jeffery1922,jeffrey1981,crowdy2011}. We state it in terms of fundamental flow singularities to connect it to the existing literature on image systems.

It is convenient to express the problem in the complex plane $z \equiv x+\ii y$, admitting a formulation of fluid dynamics in terms of analytic functions~\cite{langlois1964}. In this domain, we place two non-slip discs of radii $r_1$, $r_2$ along the imaginary axis at $z_1$ and $z_2$ respectively. The flow about these discs can be expressed elegantly as being due to flow singularities at two points $z=\pm \ii R$ inside the discs,
\begin{equation}
\begin{gathered}
\pm \frac{\tilde{\Phi}}{8 \pi \mu}\text{Sto} \left. \right|_{\mp \mathrm{i}R}
+\biggr(\mp \frac{\tilde{T}}{4 \pi \mu} + \frac{R \mathrm{Re}\, \tilde{\Phi}}{4 \pi \mu} \biggl) \text{Rot}\left. \right|_{\mp \mathrm{i}R} \\
+\biggl( \frac{R \mathrm{Im}\,\tilde{\Phi}}{4 \pi \mu} \mp \frac{\mathrm{i}\tilde{T}}{4 \pi \mu}\biggr)\frac{\kappa_{1,2}^2 }{1+\kappa_1^2 \kappa_2^2} \biggl(  \text{Str}\left. \right|_{\mp \mathrm{i}R} \pm 2\mathrm{i} R \, \text{Dip}\left. \right|_{\mp \mathrm{i}R} \biggr),
\end{gathered}
\label{eq:flow_solution}
\end{equation}
where $\text{Sto}$, $\text{Rot}$, $\text{Str}$ and $\text{Dip}$ are, respectively, a stokeslet, rotlet, stresslet and dipole, each of unit strength and located at the position denoted by the subscript, defined as in~\cite{crowdy2010a}; and $\kappa_1$, $\kappa_2$ set the disc radii and positions of their centres as
\begin{equation}
r_{1,2}=\frac{2R \kappa_{1,2}}{|1-\kappa_{1,2}^2|}, \quad z_{1,2} = \mp \mathrm{i} R\frac{1+\kappa_{1,2}^2}{1-\kappa_{1,2}^2}.
\label{a_geom}
\end{equation}
A more thorough discussion of this solution will appear elsewhere~\cite{papavassiliou2015}.

The case of two finite-sized discs is given by choosing $\kappa_1,\kappa_2<1$. However, there are two other configurations for which  eq.~(\ref{eq:flow_solution}) holds: if $\kappa_i>1$ disc $i$ is inverted and the fluid region is between two non-concentric nested circles, while if $\kappa_i=1$ then disc $i$ becomes an infinite planar boundary~\cite{jeffery1922}. In the latter case, shrinking the other disc to a point ($\kappa \to 0$) we recover the well-known image systems for stokeslets and rotlets close to an infinite no-slip wall which, remarkably, are the same as those in three dimensions~\cite{blake1974}.

Although eq.~(\ref{eq:flow_solution}) is exact, the restriction that there is no net force or torque means the reciprocal theorem can only be used to calculate the relative motions of the discs. This is easy to see from the singularity structure: the equal and opposite stokeslets and rotlets contribute forces and torques which cancel, while the second term in the rotlet opposes the couple induced by the point-force pair. This result explains why the existing solution for Stokes drag on a disc near a wall~\cite{jeffrey1981} does not exhibit the Stokes paradox: the wall exerts force and torque on the fluid equal and opposite to the dragging of the disc, regularising the asymptotic flow. The same observation is made in three dimensions by Blake \& Chwang~\cite{blake1974}.

For completeness, we give the motions of the discs under this dragging,
\begin{align}
\tilde{\Omega}_{1,2} &=\pm \frac{(1-\kappa_1^2\kappa_2^2)}{(1+\kappa_1^2\kappa_2^2) r_{1,2}^2}\frac{\tilde{T}}{4 \pi \mu }-\frac{2 R}{r_{1,2}^2} \frac{\mathrm{Re}\, \tilde{\Phi}}{8 \pi \mu }, \label{a_omega12}
\\
\mathrm{Re} \, \tilde{U}_{1,2} &=-\frac{(1-\kappa_1^2)(1-\kappa_2^2)}{2R(1+\kappa_1^2\kappa_2^2)}\frac{\tilde{T}}{4 \pi \mu }\mp \log[\kappa_{1,2}^2]\frac{\mathrm{Re} \, \tilde{\Phi}}{8 \pi \mu }, \label{a_reu}
\\
\mathrm{Im} \, \tilde{U}_{1,2}&= \mp \biggl(\frac{(1 \mp \kappa_{1}^2)(1 \pm \kappa_{2}^2)}{(1+\kappa_1^2\kappa_2^2)} + \log [\kappa_{1,2}^2]\biggr)\frac{\mathrm{Im}\, \tilde{\Phi}}{8 \pi \mu }. \label{a_imu}
\end{align}
These expressions demonstrate an exchange symmetry in the motion of the two discs: the force drives opposing translation but co-rotation, while the torque gives opposing rotations and co-translation, with equal magnitude when the discs are of equal size. They also show explicitly that as the disc radii diverge they become immobile.

\section{Application of the solution}

\begin{figure}[t]
\includegraphics[width=0.48\textwidth]{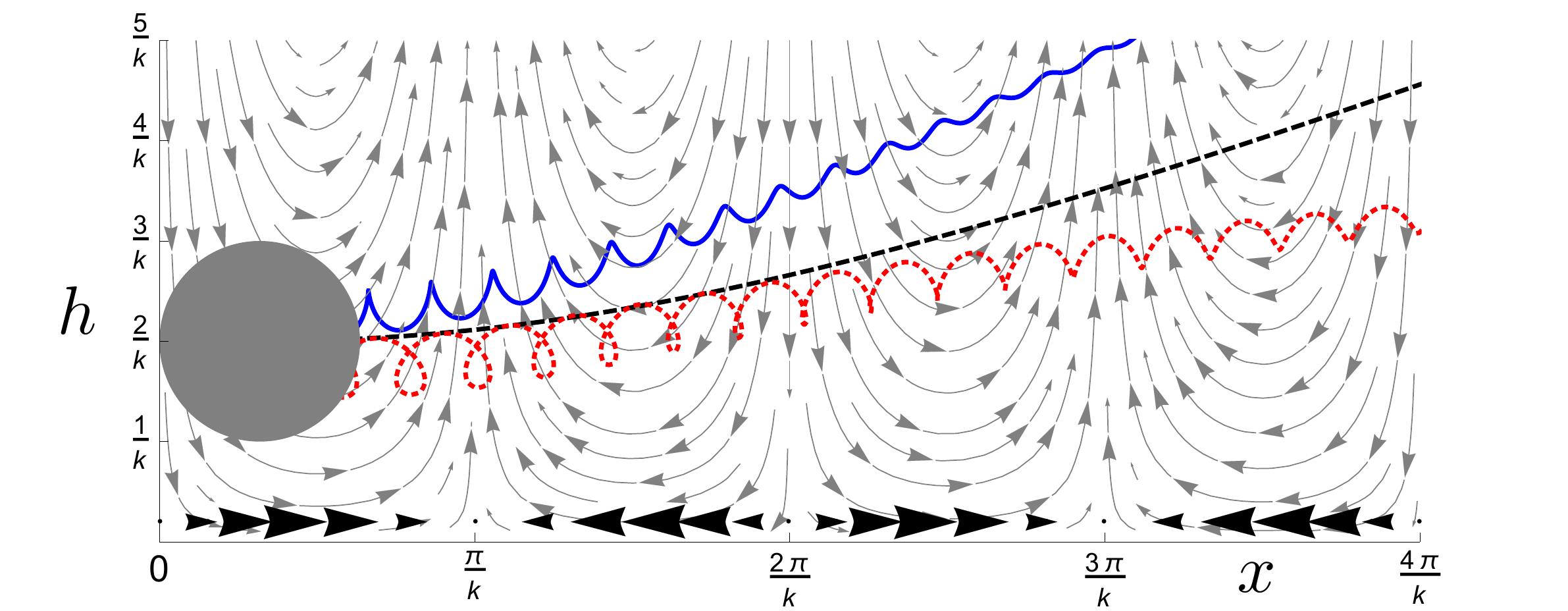}
        \caption{Flow and trajectories of a finite-sized squirmer due to an active wall in two dimensions. In the absence of wall activity the squirmer is deflected (black line). If the wall generates a metachronal wave, here with a wavelength comparable to the squirmer's size, the deflection is enhanced or suppressed respectively when the wave opposes (blue) or co-moves (red) with the swimmer. This is qualitatively similar to the three-dimensional approximate case shown in fig.~\ref{fig0b}.}
        \label{fig5b}
\end{figure}

Given the exact solution to the Stokes drag, eq.~(\ref{eq:flow_solution}), it is straightforward to construct an exact integration kernel to determine the interaction of any pair of active or inactive circular boundaries, for any slip velocities. The only limitation is that only the relative motion may be determined, since we cannot specify the forces and torques independently. If, however, one of the two boundaries corresponds to an immobile object the behaviour of the other may be determined completely.

Consider first a passive disc of radius $a$ advected by an active wall ($\kappa_1 \to 1$) with a surface metachronal wave, as before. The motion of the disc is given by $\vect{U}\cdot \tilde{\vect{\Phi}}+ \Omega T = - \int_{-\infty}^{\infty} \upd x\, U_0\sin(k x-\omega t)\hat{\vect{x}} \cdot \tilde{\tens{\sigma}} \cdot \hat{\vect{z}}$, or, explicitly,
\begin{equation}
\begin{gathered}
\mathrm{Re}\,U=U_0\mathrm{e}^{-k \sqrt{h^2-a^2}} \sin(kx-\omega t), \\
\mathrm{Im}\,U=-U_0\mathrm{e}^{-k \sqrt{h^2-a^2}}  k (h-a^2/h)\cos(k x-\omega t),\\
\Omega=U_0\mathrm{e}^{-k \sqrt{h^2-a^2}}  k\sqrt{1-a^2/h^2} \sin(k x-\omega t),
\end{gathered}
\label{activewall2d}
\end{equation}
and is broadly the same as eq.~(\ref{activewall}) in the three-dimensional case. The finite size of the tracer is encoded by the replacement of $h$ with $\sqrt{h^2-a^2}$. When $a \to 0$ we recover, as before, an exact flow field external to the wall due to the activity, shown in fig.~\ref{fig5b}.

For a squirming particle we adopt a minimal model consisting of the two squirming modes determining the swimming speed $V$ and the force dipole $W$, so that the slip velocity is given by
\begin{equation}
u_{s}=2V\sin [ \phi-\theta]+2W\sin \bigl[ 2(\phi-\theta)\bigr],
\label{6.01}
\end{equation}
at the point on the surface parametrised by the angle $\phi$ with respect to the direction of the head, $\theta$. When $V$ and $W$ have the same sign the swimmer is termed \emph{contractile}, otherwise, it is \emph{extensile}~\cite{ramaswamy2010}; these two types of activity are related by time inversion~\cite{pooley2007}. Applying the reciprocal theorem for this slip velocity gives the motion as
\begin{align}
&\begin{gathered}
\mathrm{Re}[U_2-U_1]=
(1+\kappa_2^2)(V_2\cos \theta_2-\kappa_2 W_2\sin 2\theta_2) \\
-(1+\kappa_1^2)(V_1\cos \theta_1 +\kappa_1 W_1\sin 2\theta_1),
\end{gathered}  \label{6.04a}
\\
&\begin{gathered}
\mathrm{Im}[U_2-U_1]=
\biggl [ (1-\kappa_2^2)(V_2\sin \theta_2+2\kappa_2 W_2\cos 2\theta_2) \\
 - (1-\kappa_1^2)(V_1\sin \theta_1 -2\kappa_1W_1\cos 2\theta_1) \biggr] \frac{(1-\kappa_1^2\kappa_2^2)}{(1+\kappa_1^2\kappa_2^2)} ,
 \end{gathered}  \label{6.04b}
 \\
& \begin{gathered}
\Omega_2-\Omega_1=
\Biggl [ (1-\kappa_2^2) \biggl( \frac{V_2\cos \theta_2}{(1-3\kappa_1^2 \kappa_2^2)}+ \frac{W_2\sin 2\theta_2}{\kappa_1^2\kappa_2} \biggr) +\\
 (1-\kappa_1^2) \biggl( \frac{V_1\cos \theta_1}{(1-3\kappa_1^2 \kappa_2^2)} - \frac{W_1\sin 2\theta_1}{\kappa_1\kappa_2^2}\biggr)
\Biggr ]\frac{2(1-3\kappa_1^2 \kappa_2^2)\kappa_1^2 \kappa_2^2  }{R(1+\kappa_1^2\kappa_2^2)}.
 \end{gathered}  \label{6.04c}
\end{align}
We give this expression for the general case of two arbitrary circular boundaries with slip velocities given by eq.~(\ref{6.01}), from which specific cases can be recovered by appropriate choices of $\kappa_1$, $\kappa_2$. Letting $\kappa_1 \to 1$, and $V_1,W_1 \to 0$, gives the interaction of a squirming disc with a wall~\cite{crowdy2011}. Superposition with eq.~(\ref{activewall2d}) gives the motion of a squirmer in the presence of a metachronal surface wave. The general phenomenology in this case is illustrated in fig.~\ref{fig5b}; when swimming against the metachronal wave the deflection from the surface is increased (blue line), while swimming with the metachronal wave decreases the deflection (red line). Qualitatively, this is the same as we found in three dimensions.

\begin{figure}[t]
\includegraphics[width=0.48\textwidth]{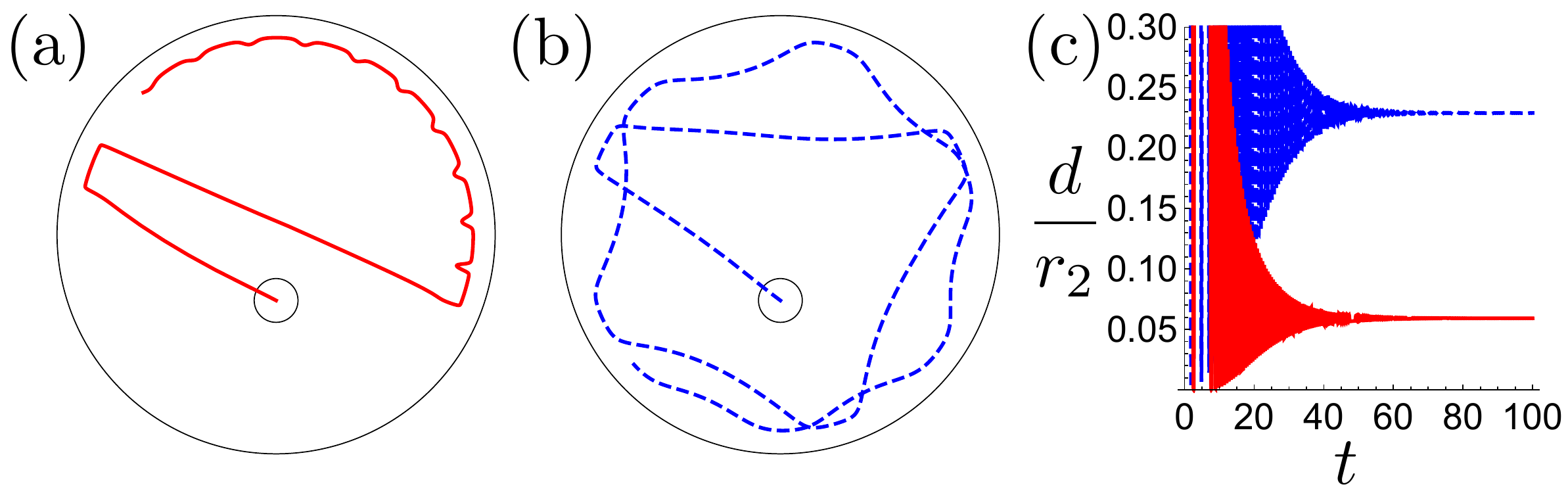}
        \caption{A squirmer of radius $r_2=0.1$ in a tank of radius $r_1=1$. (a) An extensile swimmer rapidly adheres to the wall. (b) A contractile swimmer with the same initial orientation and squirming strengths takes a longer time but also settles on a trajectory following the boundary in the opposite sense to the extensile swimmer. (c) The gap between the swimmer and the wall for the two trajectories. The asymptotic separation is significantly larger for the contractile swimmer.}
        \label{fig8}
\end{figure}

Next, consider a swimmer inside a no-slip circular tank. If $V=0$ the swimmer is apolar and we recover previous results~\cite{davis2012} (not shown). In particular, the symmetry of the problem means such a swimmer gets stuck in the centre of the tank. Including self-propulsion, the swimmer crashes into the boundary and settles into a trajectory following it, with the reorientation and sense of the trajectory determined by the microscopic details of the squirming, as shown in fig.~\ref{fig8}. For the initial conditions studied, extensile swimmers are found to be attracted more strongly to the boundary than contractile ones.

We now turn to interactions between two finite-sized squirmers. The fact that we can only find relative translation and rotation becomes problematic in determining the motion; in the previous cases a frame of reference was defined by the wall or tank. There are two resolutions: to make one disc an immobile, passive `post' which fixes a reference frame as before; or to make both squirmers identical so that each contributes equally to the motion in some centre-of-mass frame, which cannot be determined.

In the former case, taking $V_1,W_1\to 0$ disc 1 becomes a no-slip post. Then we can transform eqs.~(\ref{6.04a})-(\ref{6.04c}) to polar coordinates $(d,\varphi)$ centred on the post. By the rotational symmetry about the post, this dynamical system depends only on the relative angle $\theta_R \equiv \theta-\varphi$. This allows us to form a lower-dimensional system, $\bigl(d(t),\theta_R(t)\bigr)$, integrable by separation of variables. A fixed point in this phase space giving a bound trajectory has previously been seen for the interaction of a swimmer with an infinite wall~\cite{crowdy2011,or2011}. We find that this effect persists when the post is finite in size, corresponding to a bound orbit of the post by the swimmer. As the size of the swimmer approaches the size of the post, the location of this orbit diverges algebraically as $d \sim (1-r_2/r_1)^{-1/2}$. When the swimmer is larger than the post we see generic deflection.

\begin{figure}[t]
\includegraphics[width=0.48\textwidth]{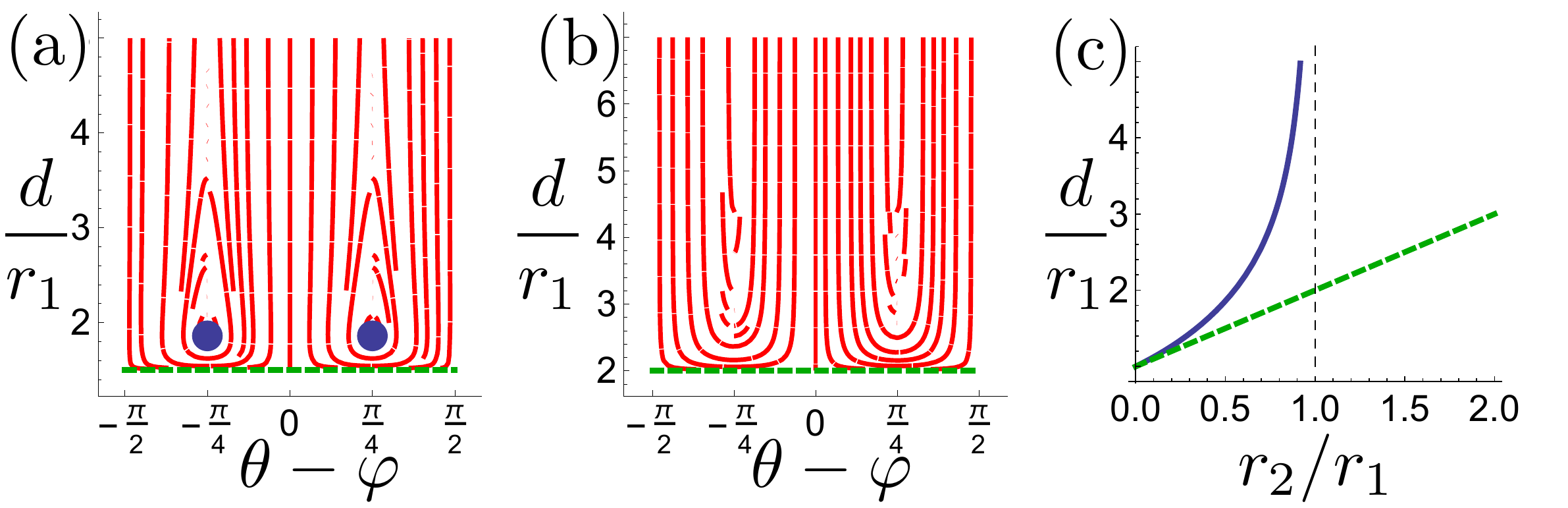}
        \caption{(a) Phase flow for a swimmer with radius 0.5 relative to the post. (b) Phase flow for a swimmer with equal radius to the post. When the swimmer radius is smaller than the post radius there are closed trajectories about a circular orbit (blue). As the swimmer radius approaches the post radius from below the distance of this circular orbit diverges as $(1-r_2/r_1)^{-1/2}$. The exclusion is shown in green. (c) Existence of the circular orbit (blue) compared to the excluded radius (green).}
        \label{fig6}
\end{figure}

\begin{figure}[t]
\includegraphics[width=0.48\textwidth]{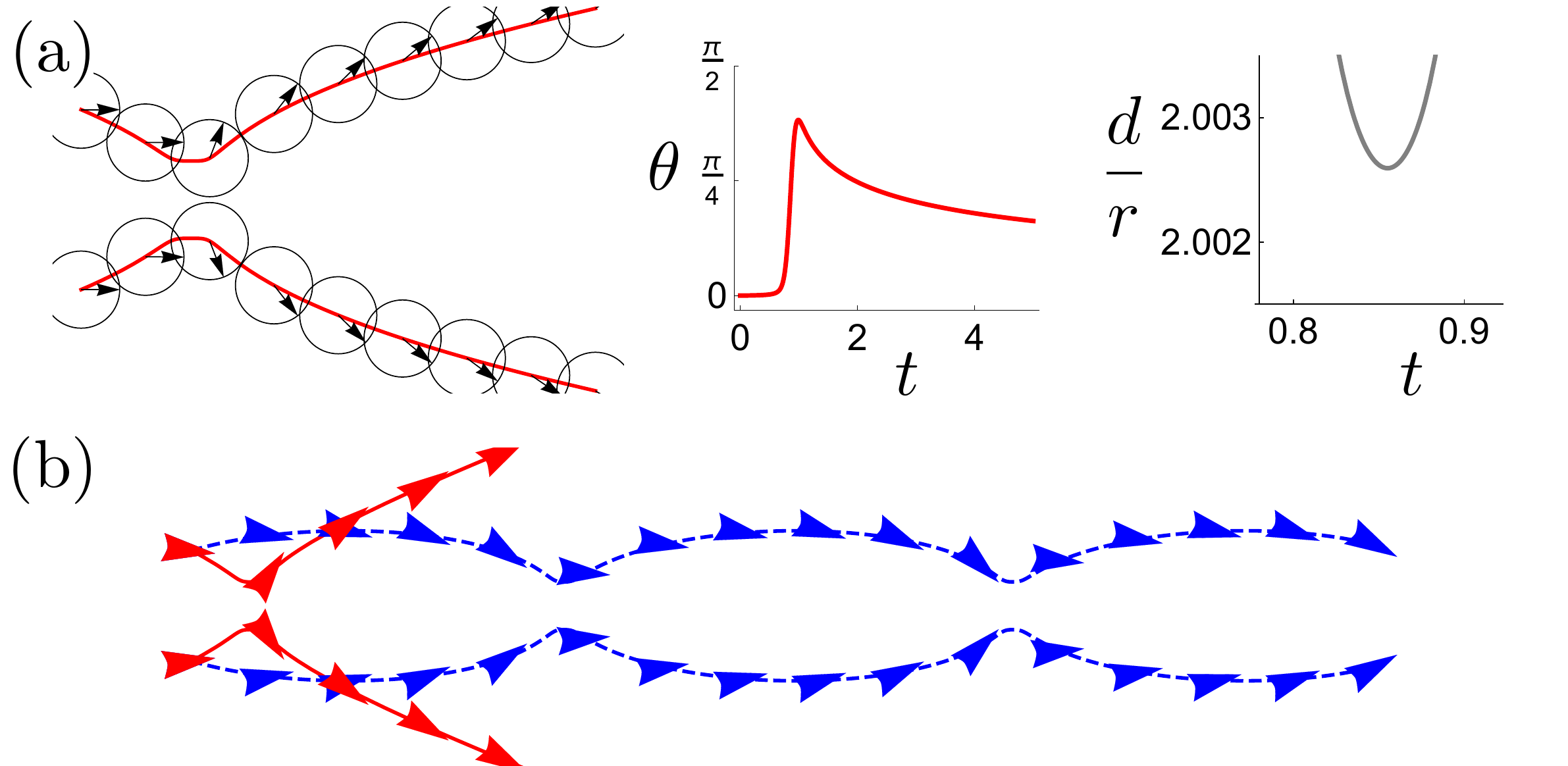}
        \caption{Interactions of two identical squirmers in the centre-of-mass frame, swept out in time. (a) Two initially parallel extensile swimmers are attracted and undergo the reorientation shown by the red trace. The minimum separation during the collision is shown by the grey curve. (b) Trajectory of two extensile (red) and contractile (blue) swimmers with an initial tilt towards each other. Contractile swimmers exhibit a periodic bound trajectory. By the time-reversal symmetry of contractile and extensile activity, this is seen in extensile swimmers when the initial orientation is outwards.}
        \label{fig7}
\end{figure}

Finally, consider the hydrodynamic interaction between two identical squirmers and an initial condition with a mirror symmetry. Then each will rotate in an opposite sense with respect to the centre of mass, with half the magnitude of $\Omega_2-\Omega_1$. 
We illustrate in fig.~\ref{fig7} some of the close proximity interactions, determined by exact hydrodynamics. The reorientation and distance of closest approach for a typical collision is shown in fig.~\ref{fig7}(a), where the squirmers start exactly parallel. Angling two contractile swimmers slightly towards each other can lead to bound trajectories, shown in blue in fig.~\ref{fig7}(b). This occurs for extensile swimmers when they initially point away from each other, as a consequence of the time-reversal symmetry between the two types~\cite{pooley2007}.

\section{Discussion}

The reciprocal theorem is a standard technique for understanding swimmer motion~\cite{stone1996,squires2004,golestanian2007,crowdy2011,lauga2014,masoud2014}. The reinterpretation presented here shows how it can be also used to obtain all aspects of swimmer hydrodynamics. We gave simple illustrations in three dimensions for the motion of a squirmer in the presence of a planar boundary, reproducing literature results without approximation of the swimming strategy, and extending to interactions with a ciliated surface. Adopting standard exact results for Stokes flows in two dimensions~\cite{jeffery1922} allowed for a wide range of swimmer hydrodynamics and near-field interactions to be solved. The three-dimensional results are in good qualitative agreement with these exact results, suggesting that one may have some confidence in their validity, even close to surfaces.

If the stress tensor of a Stokes drag problem is known, the corresponding swimmer hydrodynamics is also given by the reciprocal theorem. This allows a large number of swimmer hydrodynamics problems to be solved exactly in both two and three dimensions, by adoption of literature results~\cite{papavassiliou2015}. Another interesting generalisation would be to incorporate free surfaces with boundary conditions of specified stress rather than slip velocity~\cite{kanevsky2010,masoud2014}. 

The reciprocal theorem gives swimmer hydrodynamics in terms of an integral against the boundary data, with the integration kernel identified as the stress tensor of a conjugate Stokes drag problem. This is analogous to, but distinct from, boundary integral methods~\cite{pozrikidis1992} for hydrodynamics, where the flow is given by integrating a force distribution over the boundaries. Here, the boundary conditions are the slip velocities directly, and the integration kernel is the stress tensor for a conjugate Stokes drag problem. Because the stress tensor decays more rapidly than the Green function for a point force, this approach may yield better convergence results in general.

\acknowledgments
We thank Matteo Contino, Darren Crowdy, Eric Lauga, Tom Machon and Marco Polin for fruitful discussion and insight. This work was partially supported by the UK EPSRC through grant no.\ A.MACX.0002.

\bibliographystyle{eplbib}
\renewcommand{\bibname}{References}
\bibliography{DPGPA1503_2}

\end{document}